\documentclass[twocolumn,preprintnumbers,amsmath,amssymb,superscriptaddress]{revtex4}

\usepackage{amsmath,amsfonts,amssymb}
\usepackage[english]{babel}
\usepackage[latin1]{inputenc}
\usepackage[T1]{fontenc}
\usepackage{color}
\usepackage{float}
\usepackage{verbatim}
\usepackage{graphicx}
\usepackage{bm}
\usepackage{mathtools}
\usepackage{stmaryrd}
\usepackage{anyfontsize}

\usepackage{color}

\usepackage{bibunits}

\begin{document}

\author{Justin D. Yeakel} \affiliation{School of Natural Sciences, University
  of California, Merced, Merced, CA 95340, USA}

\author{Christopher P. Kempes} \affiliation{The Santa Fe Institute, 1399 Hyde
  Park Road, Santa Fe, NM 87501, USA}

\author{Sidney Redner} \affiliation{The Santa Fe Institute, 1399 Hyde Park
  Road, Santa Fe, NM 87501, USA}

\title{The dynamics of starvation and recovery}

\begin{abstract} 
The eco-evolutionary dynamics of species are fundamentally linked to the energetic constraints of its constituent individuals. Of particular importance is the interplay between reproduction and the dynamics of starvation and recovery. To elucidate this interplay, we introduce a nutritional state-structured model that incorporates two classes of consumer: nutritionally replete, reproducing consumers, and undernourished, non-reproducing consumers. We obtain strong constraints on starvation and recovery rates by deriving allometric scaling relationships and find that population dynamics are typically driven to a steady state. Moreover, these rates fall within a `refuge' in parameter space, where the probability of population extinction is minimized. We also show that our model provides a natural framework to predict maximum mammalian body size by determining the relative stability of an otherwise homogeneous population to a competing population with altered percent body fat. This framework provides a principled mechanism for a selective driver of Cope's rule.
\end{abstract}

\maketitle

\begin{bibunit}[unsrt]

  The behavioral ecology of all organisms is influenced by their energetic
  states, which directly impacts how they invest reserves in uncertain
  environments.  Such behaviors are generally manifested as tradeoffs between
  investing in somatic maintenance and growth, or allocating energy towards
  reproduction~\citep{Martin:1987dl,Kirk:1997cc,Kempes:2012hy}.  The timing of
  these behaviors responds to selective pressure, as the choice of the
  investment impacts future
  fitness~\citep{Mangel:1988uaa,Mangel:2014kz,Yeakel:2013hi}.  The influence of
  resource limitation on an organism's ability to maintain its nutritional
  stores may lead to repeated delays or shifts in reproduction over the course
  of an organism's life.

  The balance between (a) somatic growth and maintenance, and (b) reproduction depends on resource availability~\citep{Morris:1987eo}.
  For example, reindeer invest less in calves born after harsh winters (when the mother's energetic state is depleted) than in calves born after moderate winters~\citep{Tveraa:2003fq}.
  Many bird species invest differently in broods during periods of resource scarcity~\citep{Daan:1988va,Jacot:2009dv}, sometimes delaying or even foregoing reproduction for a breeding season~\citep{Martin:1987dl,Stearns:1989ip,Barboza:2002in}.
  Even freshwater and marine zooplankton have been observed to avoid reproduction under nutritional stress~\citep{Threlkeld:1976ih}, and those that do reproduce have lower survival rates~\citep{Kirk:1997cc}.
  Organisms may also separate maintenance and growth from reproduction over space and time: many salmonids, birds, and some mammals return to migratory breeding grounds to reproduce after one or multiple seasons in resource-rich environments where they accumulate reserves~\citep{Weber:1998jg,Mduma:1999cp,Moore:2014hi}.

  Physiology also plays an important role in regulating reproductive expenditures during periods of resource limitation.
  Many mammals (47 species in 10 families) exhibit delayed implantation, whereby females postpone fetal development until nutritional reserves can be accumulated~\citep{Mead:1989dt,Sandell:1990kw}.
  Many other species (including humans) suffer irregular menstrual cycling and higher abortion rates during periods of nutritional stress~\citep{Bulik:1999eo,Trites:2003cc}.
  In the extreme case of unicellular organisms, nutrition directly controls growth to a reproductive state \citep{Glazier:2009hq,Kempes:2012hy}. The existence of so many independently evolved mechanisms across such a diverse suite of organisms highlights the near-universality of the fundamental tradeoff between somatic and reproductive investment.

  Including individual energetic dynamics~\citep{Kooi2000} in a
  population-level framework~\citep{Kooi2000,Sousa:2010ez} is
  challenging~\citep{Diekmann:2010da}.  A common simplifying approach is the
  classic Lotka-Volterra (LV) model, which assumes that consumer population
  growth rate depends linearly on resource density~\citep{murdoch:2003}. Here,
  we introduce an alternative approach---the Nutritional State-structured Model
  (NSM)---that accounts for resource limitation via explicit starvation. In
  contrast to the LV model, the NSM incorporates two consumer states: hungry
  and full, with only the former susceptible to mortality and only the latter
  possessing sufficient energetic reserves to reproduce.  Additionally, we
  incorporate allometrically derived constraints on the time scales for
  reproduction~\citep{Kempes:2012hy}, starvation, and recovery.  Our model
  makes several basic predictions: (i) the dynamics are typically driven to a
  refuge far from cyclic behavior and extinction risk, (ii) the steady-state
  conditions of the NSM accurately predict the measured biomass densities for
  mammals described by Damuth's law~\citep{Damuth:1987kr,allen2002,enquist1998,Pedersen:2017he},
  (iii) there is an allometrically constrained upper-bound for mammalian body size, and
  (iv) the NSM provides a selective mechanism for the evolution of larger body size, known as Cope's rule~\citep{Alroy:1998p1594,Clauset:2009fh,Smith:2010p3442,Saarinen:2014br}.



  \noindent \paragraph*{{\bf Nutritional state-structured model (NSM).}}
  We begin by defining the nutritional state-structured population model, where the consumer population is partitioned into two states: (a) an
  energetically replete (full) state $F$, where the consumer reproduces at a
  constant rate $\lambda$ and does not die from starvation, and (b) an
  energetically deficient (hungry) state $H$, where the consumer does not
  reproduce but dies by starvation at rate $\mu$. The dynamics of the
  underlying resource $R$ are governed by logistic growth with an intrinsic
  growth rate $\alpha$ and a carrying capacity $C$. The rate at which consumers
  transition between states and consume resources is dependent on their number,
  the abundance of resources, the efficiency of converting resources into
  metabolism, and how that metabolism is partitioned between maintenance and
  growth purposes.  We provide a physiologically and energetically mechanistic
  model for each of these dynamics and constants (see the Supplementary Information
  (SI)), and show that the system produces a simple non-dimensional form which
  we describe below.

  \begin{figure}
  \centering
  \includegraphics[width=0.45\textwidth]{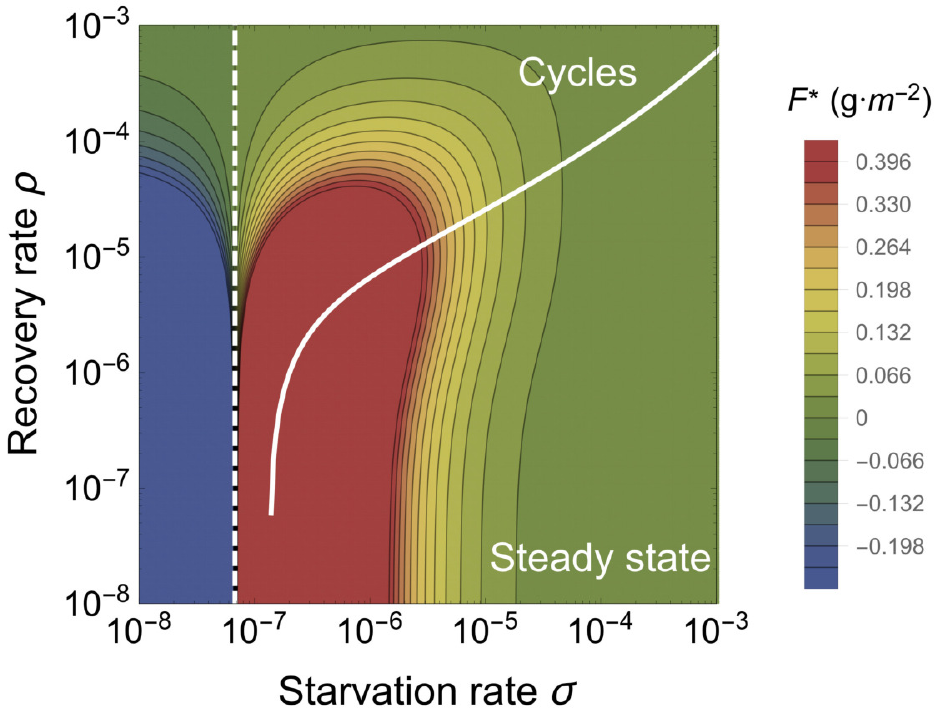}
  \caption{\small{ The transcritical (TC; dashed line) and Hopf bifurcation (solid line) as a
    function of the starvation rate $\sigma$ and recovery rate $\rho$ for a 100g consumer.  These
    bifurcation conditions separate parameter space into unphysical (left of the TC), cyclic,
    and steady state dynamic regimes.  The colors show the steady state densities for the energetically replete consumers $F^*$.  
  }\label{fig:fp}}
  \end{figure}

  Consumers transition from the full state $F$ to the hungry state $H$ at a
  rate $\sigma$---the starvation rate---and also in proportion to the absence
  of resources $(1-R)$ (the maximum resource density has been non
  dimensionalized to 1; see SI).  Conversely, consumers recover from state $H$
  to state $F$ at rate $\xi \rho$ and in proportion to $R$, where $\xi$
  represents a ratio between maximal resource consumption and the carrying
  capacity of the resource. 
  The resources that are eaten by hungry consumers (at rate $\rho R + \delta$)
  account for their somatic growth ($\rho R$) and maintenance ($\delta$).  Full
  consumers eat resources at a constant rate $\beta$ that accounts for maximal
  maintenance and somatic growth (see the SI for mechanistic derivations of
  these rates from resource energetics).
  The NSM represents an ecologically motivated fundamental extension of the
  idealized starving random walk model of foraging, which focuses on resource
  depletion, to include reproduction and resource
  replenishment~\citep{Benichou:2014wu,Benichou:2016wl,Chupeau:2016jf}, and is
  a more general formulation than previous models that incorporate
  starvation~\citep{Persson:1998hz}.

  In the mean-field approximation, in which the consumers and resources are
  perfectly mixed, their densities are governed by the rate equations

  \begin{align}
  \label{eq:system}
  \begin{split}
  \dot{F} &= \lambda F + \xi \rho RH - \sigma \left(1-R\right)F,  \\
  \dot{H} &= \sigma \left(1-R\right)F - \xi \rho RH - \mu H,  \\
  \dot{R} &= \alpha \left(1-R\right)R -\left(\rho R+\delta\right)H-\beta F.
  \end{split}
  \end{align}

  This system of nondimensional equations follows from a set of first-principle
  relationships for resource consumption and growth (see the SI for a full derivation and the dimensional form).
  Notice that the total consumer density $F+H$ evolves according to $\dot{F}+\dot{H}=\lambda F-\mu H$.
  This resembles the equation of motion for the predator density in the LV model~\citep{murray2011mathematical}, except that the resource density does not appear in the growth term.
  The rate of reproduction is independent of resource density because the full
  consumer partitions a constant amount of energy towards reproduction, whereas
  a hungry consumer partitions no energy towards reproduction.  Similarly, the
  consumer maintenance terms ($\delta H$ and $\beta F$) are also independent of
  resource density because they represent a minimal energetic requirement for
  consumers in the $H$ and $F$ state, respectively.


  \noindent \paragraph*{ {\bf Steady states of the NSM.}} From the single
  internal fixed point (Eq.~\eqref{eq:ss}, see Methods), an obvious constraint
  on the NSM is that the reproduction rate $\lambda$ must be less than the
  starvation rate $\sigma$, so that the consumer and resource densities are
  positive.
  The condition $\sigma = \lambda$ represents a transcritical (TC)
  bifurcation~\citep{Strogatz:2008wo} that demarcates a physical from an
  unphysical (negative steady-state densities) regime.  The biological
  implication of the constraint $\lambda<\sigma$ has a simple
  interpretation---the rate at which a macroscopic organism loses mass due to
  lack of resources is generally much faster than the rate of reproduction.  As
  we will discuss below, this inequality is also a natural consequence of
  allometric constraints~\citep{Kempes:2012hy} for organisms within empirically
  observed body size ranges. 

  In the physical regime of $\lambda<\sigma$, the fixed point \eqref{eq:ss} may either be a stable node or a limit cycle (Fig.~\ref{fig:fp}).
  In continuous-time systems, a limit cycle arises when a pair of complex conjugate eigenvalues crosses the imaginary axis to attain positive real parts~\citep{GuckHolmes}.
  This Hopf bifurcation is defined by ${\rm Det}({\bf S}) = 0$, with $\bf S$ the Sylvester matrix, which is composed of the coefficients of the characteristic polynomial of the Jacobian matrix~\citep{Gross:2004p2428}.
  As the system parameters are tuned to be within the stable regime, but close to the Hopf bifurcation, the amplitude of the transient cycles becomes large.
  Given that ecological systems are constantly being perturbed~\citep{Hastings:2001jh}, the onset of transient cycles, even though they decay with time in the mean-field description, can increase extinction risk~\citep{Neubert:1997wk,Caswell:2005eo,Neubert:2009td}.

  When the starvation rate $\sigma\gg\lambda$, a substantial fraction of the
  consumers are driven to the hungry non-reproducing state.  Because
  reproduction is inhibited, there is a low steady-state consumer density and a
  high steady-state resource density.  However, if $\sigma/\lambda\to 1$ from
  above, the population is overloaded with energetically-replete (reproducing)
  individuals, thereby promoting transient oscillations between the consumer
  and resource densities (Fig.~\ref{fig:fp}).  If the starvation rate is low
  enough that the Hopf bifurcation is crossed, these oscillations become
  stable.  This threshold occurs at higher values of the starvation rate as the recovery rate $\rho$ increases, such that the range of parameter space giving rise to cyclic dynamics also increases with higher recovery rates.\\


  \begin{figure}
  \centering
  \includegraphics[width=0.4\textwidth]{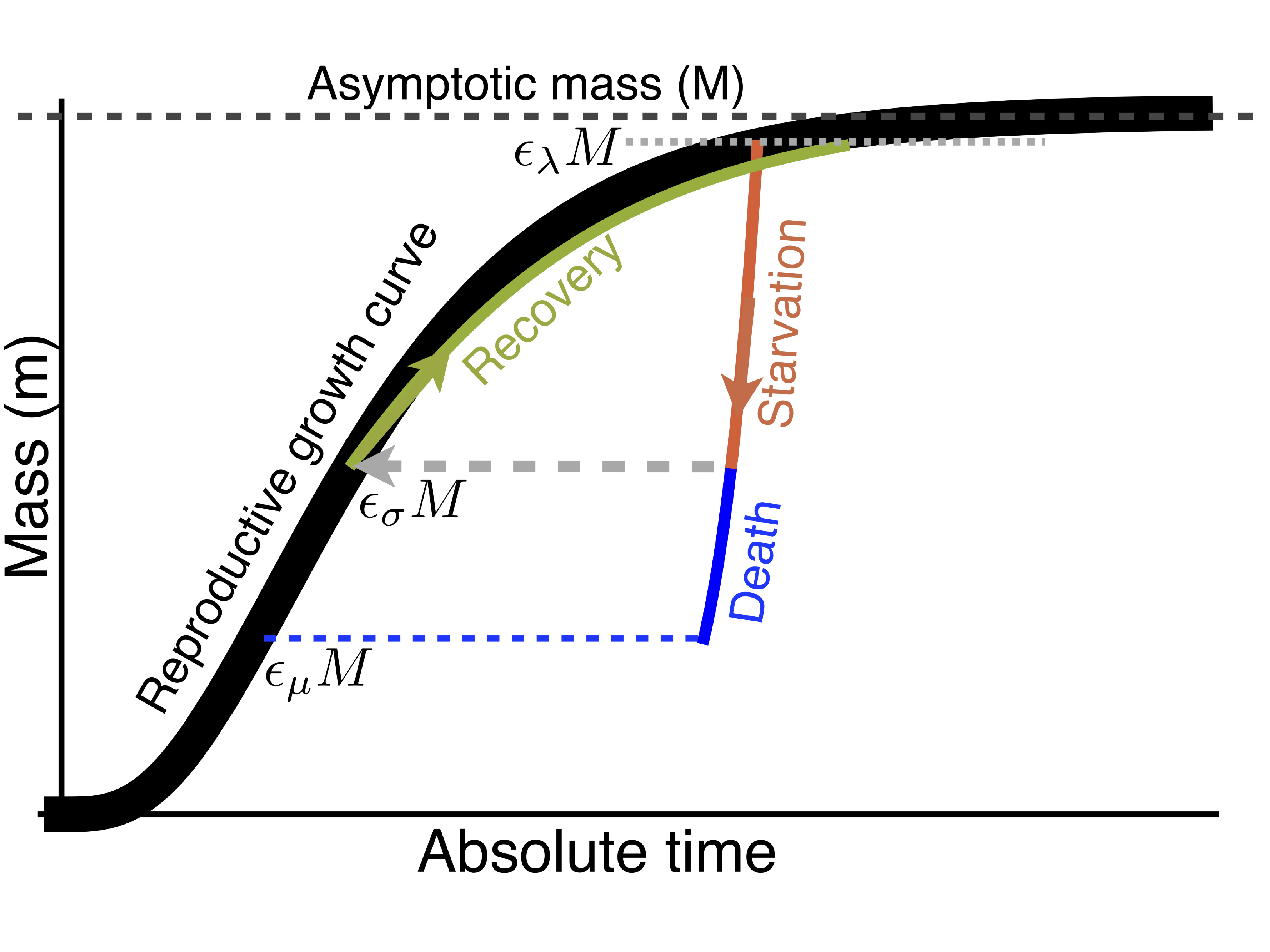}
  \caption{\small{ The growth trajectory over absolute time of an individual organism as a function of body mass.  
  Initial growth follows the black trajectory to an energetically replete reproductive adult mass of $m=\epsilon_\lambda M$ (see Methods). 
  Starvation follows the red trajectory to $m = \epsilon_\sigma \epsilon_\lambda  M$. 
  Recovery follows the green curve to the replete adult mass, where this trajectory differs from the original growth because only fat is being regrown which requires a longer time to reach $\epsilon_\lambda M$. 
  Alternatively, death from starvation follows the blue trajectory to $m=\epsilon_\mu \epsilon_\lambda  M$.}\label{fig:growth}}
  \end{figure}

  \noindent {\bf Results}
  \noindent \paragraph*{{\bf The allometry of extinction risk.}} While there
  are no {\it a priori} constraints on the parameters in the NSM, we expect
  that each species should be restricted to a distinct portion of the parameter
  space.  We use allometric scaling relations to constrain the covariation of
  rates in a principled and biologically meaningful manner (see Methods).
  Allometric scaling relations highlight common constraints and average trends
  across large ranges in body size and species diversity. Many of these
  relations can be derived from a small set of assumptions.  In the Methods we
  describe our framework to determine the covariation of timescales and rates
  across a range of body sizes for each of the key parameters of our model
  (cf.\ Ref.~\citep{Yodzis:1992hg}).

  Nearly all of the rates described in the NSM are determined by consumer
  metabolism, which can be used to describe a variety of organismal features
  \citep{Brown:2004wq}.  We derive, from first principles, the relationships
  for the rates of reproduction, starvation, recovery, and mortality as a
  function of an organism's body size and metabolic rate (see Methods).
  Because we aim to explore the starvation-recovery dynamics as a function of
  an organism's body mass $M$, we parameterize these rates in terms of the
  \emph{percent} gain and loss of the asymptotic (maximum) body mass,
  $\epsilon M$, where different values of $\epsilon$ define different states of
  the consumer (Fig.~\ref{fig:growth}; see Methods for derivations of
  allometrically constrained rate equations).  Although the rate equations
  \eqref{eq:system} are general and can in principle be used to explore the
  starvation recovery dynamics for most organisms, here we focus on allometric
  relationships for terrestrial-bound lower-trophic level endotherms (see the
  SI for values), specifically herbivorous mammals, which range from a minimum
  of $M\approx1$g (the Etruscan shrew \emph{Suncus etruscus}) to a maximum of
  $M\approx10^7$g (the early Oligocene Indricotheriinae and the Miocene
  Deinotheriinae).  Investigating other classes of organisms would simply
  involve altering the metabolic exponents and scalings associated with
  $\epsilon$. Moreover, we emphasize that our allometric equations (see
  Methods) describe mean relationships, and do not account for the (sometimes
  considerable) variance associated with individual species.  We note that
  including additional allometrically-scaled mortality terms to both $F$ and
  $H$ does not change the form of our model nor impact our quantitative findings
  (see SI for the derivation).

  \begin{figure}
  \centering
  \includegraphics[width=0.4\textwidth]{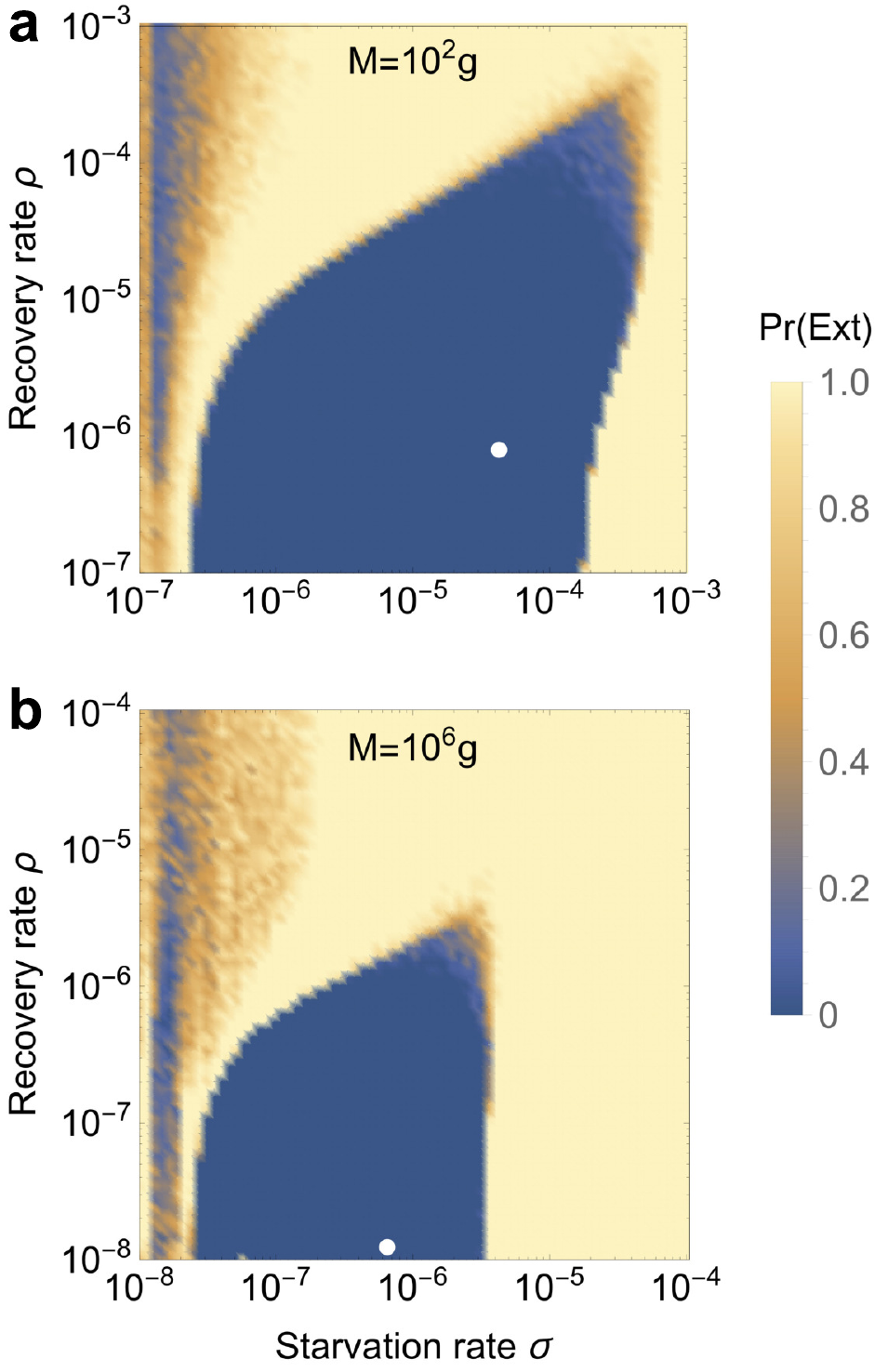}
  \caption{\small{ Probability of extinction for a consumer with ({\bf a}) $M=10^2$g and ({\bf b}) $M=10^6$g as a function of the starvation rate $\sigma$ and recovery rate $\rho$, where the initial density is given as $(XF^*,XH^*,R^*)$, where $X$ is a random uniform variable in $[0,2]$. Note the change in scale in panel {\bf b}.  Extinction is defined as the population trajectory falling below $0.2\times$ the allometrically constrained steady state. The white points denote the allometrically constrained starvation and recovery rate.}\label{fig:ext}}
  \end{figure}

  As the allometric derivations of the NSM rate laws reveal (see Methods),
  starvation and recovery rates are not independent parameters, and the
  biologically relevant portion of the phase space shown in Fig.~\ref{fig:fp}
  is constrained via covarying parameters.  Given the parameters of terrestrial
  endotherms, we find that the starvation rate $\sigma$ and the recovery rate
  $\rho$ are constrained to lie within a small region of potential values for
  the known range of body sizes $M$.  Indeed, starvation and recovery rates
  across all values of $M$ fall squarely in the steady-state region at some
  distance from the Hopf bifurcation.  This suggests that cyclic population
  dynamics should be rare, particularly in resource-limited environments.

  Higher rates of starvation result in a larger flux of the population to the hungry state.
  In this state, reproduction is absent, thus increasing the likelihood of extinction.  From the perspective of population survival, it is the rate of starvation relative to the rate of recovery that determines the long-term dynamics of the various species (Fig.~\ref{fig:fp}).
  We therefore examine the competing effects of cyclic dynamics vs.\ changes in steady-state density on extinction risk, both as functions of $\sigma$ and $\rho$.
  To this end, we computed the probability of extinction, where we define extinction as a population trajectory falling below one fifth of the allometrically constrained steady state at any time between $t=10^8$ and $t=10^{10}$.
  This procedure was repeated for 50 replicates of the continuous-time system shown in Eq.~\ref{eq:system} for organisms with mass ranging from $10^2$ to $10^6$ grams.
  In each replicate the initial densities were chosen to be $(XF^*,XH^*,R^*)$,
  with $X$ a random variable uniformly distributed in $[0,2]$.  By allowing the
  rate of starvation to vary, we assessed extinction risk across a range of
  values for $\sigma$ and $\rho$ between ca.\ $10^{-8}$ to
  $10^{-3}$. 
  Higher rates of extinction correspond to both large $\sigma$ if $\rho$ is
  small, and large $\rho$ if $\sigma$ is small.  In the former case, increased
  extinction risk arises because of the decrease in the steady-state consumer
  population density (Figs. \ref{fig:fp}b, \ref{fig:ext}).  In the latter case,
  the increased extinction risk results from higher-amplitude transient cycles
  as the system nears the Hopf bifurcation (Fig.~\ref{fig:ext}).  This
  interplay creates an `extinction refuge', such that for a constrained range
  of $\sigma$ and $\rho$, extinction probabilities are minimized.

  We find that the allometrically constrained values of $\sigma$ and $\rho$,
  each representing different trajectories along the ontogenetic curve
  (Fig. \ref{fig:growth}), fall squarely within the extinction refuge across a
  range of $M$ (Fig. \ref{fig:ext}a,b, white points). These values are close
  enough to the Hopf bifurcation to avoid low steady-state densities, yet
  distant enough to avoid large-amplitude transient cycles.  Allometric values
  of $\sigma$ and $\rho$ fall within this relatively small window, which
  supports the possibility that a selective mechanism has constrained the
  physiological conditions driving starvation and recovery rates within
  populations.  Such a mechanism would select for organism physiology that
  generates appropriate $\sigma$ and $\rho$ values that minimize extinction
  risk.  This selection could occur via the tuning of body fat percentages,
  metabolic rates, and/or biomass maintenance efficiencies.  We also find that
  as body size increases, the size of the low extinction-risk parameter space
  shrinks (Fig.~\ref{fig:ext}b), suggesting that the population dynamics for
  larger organisms are more sensitive to variability in physiological rates.
  This finding is in accordance with, and may serve as contributing support for, observations of increased extinction risk among larger mammals \cite{Liow:2008jx}.\\



\begin{figure}
  \centering
  \includegraphics[width=0.4\textwidth]{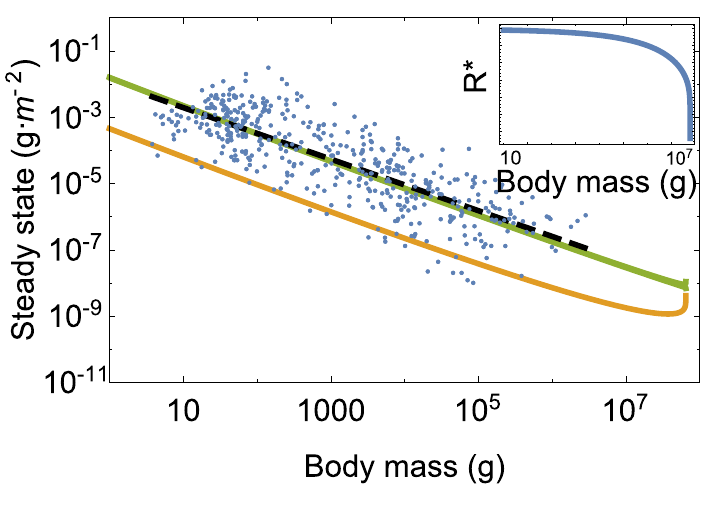}
  \caption{\small{Consumer steady states $F^*$ (green) and $H^*$ (orange) as a function of
    body mass along with the data from Damuth \citep{Damuth:1987kr}. Inset: Resource steady state $R^*$ as a function of consumer body mass.}\label{fig:mass}}
  \end{figure}

  \noindent \paragraph*{{\bf Damuth's Law and body size limits.}} The NSM
  correctly predicts that smaller species have larger steady-state population
  densities (Fig.~\ref{fig:mass}).  Similar predictions have been made for
  carnivore populations using alternative consumer-resource models
  \citep{DeLong:2012kw}.  Moreover, we show that the NSM provides independent
  theoretical support for Damuth's Law
  \citep{Damuth:1987kr,allen2002,enquist1998,Pedersen:2017he}.  Damuth's
  law shows that species abundances, $N^{*}$, follow $N^*=0.01
  M^{-0.78}$ (g m$^{-2}$). Figure \ref{fig:mass} shows that both $F^{*}$ and $H^{*}$ scale
  as $M^{-\eta}$, with $\eta\approx 3/4$,  over a wide range of organismal sizes and that $F^{*}+H^{*}$
  closely matches the best fit to Damuth's data.  Remarkably, this result
  illustrates that the steady state values of the NSM combined with the derived
  timescales naturally give rise to Damuth's law. While the previous metabolic
  studies supporting Damuth's law provided arguments for the value of the
  exponent~\citep{allen2002}, these studies are only able to infer the
  normalization constant ($0.01$ g$^{1.78}$ m$^{-2}$ in the above equation) from the data (see SI for a discussion of the energy
  equivalence hypothesis related to these metabolic arguments). Our model
  predicts not only the exponent but also the normalization constant by
  explicitly including the resource dynamics and the parameters that determine
  growth and consumption. It should be noted that density relationships of
  individual clades follow a more shallow scaling relationship than predicted
  by Damuth's law~\cite{Pedersen:2017he}.  In the context of our model,
  this finding suggests that future work may be able to anticipate these shifts
  by accounting for differences in the physiological parameters associated with
  each clade.

  With respect to predicted steady state densities, the total metabolic rate of
  $F$ and $H$ becomes infinite at a finite mass, and occurs at the same scale
  where the steady state resources vanish (Fig.~\ref{fig:mass}). This
  asymptotic behavior is governed by body sizes at which $\epsilon_{\mu}$ and
  $\epsilon_{\lambda}$ (see Fig.~\ref{fig:growth}) equal zero, causing the
  timescales (Eqn. \ref{t1}) to become infinite and the rates $\mu$ and $\lambda$ to equal
  zero.
  The $\mu=0$ asymptote occurs first when
  $f_{0}M^{\gamma-1}+u_{0}M^{\zeta-1}=1$, and corresponds to
  $(F^*,H^*,R^*)=(0,0,0)$.  This point predicts an upper bound on mammalian
  body size at $M_{\rm max}=6.54\times10^7$ (g).  Moreover, $M_{\rm max}$,
  which is entirely determined by the population-level consequences of
  energetic constraints, is within an order of magnitude of the maximum body
  size observed in the North American mammalian fossil
  record~\citep{Alroy:1998p1594}, as well as the mass predicted from an
  evolutionary model of body size evolution~\citep{Clauset:2009fh}.  We
  emphasize that the asymptotic behavior and predicted upper bound depend only
  on the scaling of body composition and are independent of the resource
  parameters.  The prediction of an asymptotic limit on mammalian size
  parallels work on microbial life where an upper and lower bound on bacterial
  size, and an upper bound on single cell eukaryotic size, is predicted from
  similar growth and energetic scaling
  relationships~\citep{Kempes:2012hy,Kempes:2016}.  It has also been shown that
  models that incorporate the allometry of hunting and resting combined with
  foraging time predicts a maximum carnivore size between $7\times10^{5}$ and
  $1.1\times10^{6}$ (g) \cite{Carbone:1999ju,Carbone:2007dz}.  Similarly, the
  maximum body size within a particular lineage has been shown to scale with
  the metabolic normalization constant
  \cite{Okie:2013ju}. 
  This complementary approach is based on the balance between growth and
  mortality, and suggests that future connections between the scaling of fat
  and muscle mass should systematically be connected with $B_{0}$ when
  comparing
  lineages. 

  \noindent \paragraph*{{\bf A mechanism for Cope's rule}} Metabolite transport
  constraints are widely thought to place strict boundaries on biological
  scaling~\citep{Brown:1993p708,West:1997cg,Brown:2004wq} and thereby lead to
  specific predictions on the minimum possible body size for
  organisms~\citep{West:2002ud}.  Above this bound, a number of energetic and
  evolutionary mechanisms have been explored to assess the costs and benefits
  associated with larger body masses, particularly for mammals.  One important
  such example is the \emph{fasting endurance hypothesis}, which contends that
  larger body size, with consequent lower metabolic rates and increased ability
  to maintain more endogenous energetic reserves, may buffer organisms against
  environmental fluctuations in resource availability~\citep{Millar:1990p923}.
  Over evolutionary time, terrestrial mammalian lineages show a significant
  trend towards larger body size---Cope's
  rule~\citep{Alroy:1998p1594,Clauset:2009fh,Smith:2010p3442,Saarinen:2014br}.
  It is thought that within-lineage drivers generate selection towards an
  optimal upper bound of roughly $10^7$ (g)~\citep{Alroy:1998p1594}, a value
  that is likely limited by higher extinction risk for large taxa over longer
  timescales~\citep{Clauset:2009fh}.  These trends are thought to be driven by
  a combination of climate change and niche
  availability~\citep{Saarinen:2014br}; however the underpinning energetic
  costs and benefits of larger body sizes, and how they influence dynamics over
  ecological timescales, have not been explored.

  The NSM predicts that the steady state resource density $R^{*}$ decreases
  with increasing body size of the consumer population (Fig.~\ref{fig:mass},
  inset), and classic resource competition theory predicts that the species
  surviving on the lowest resource abundance will outcompete others
  \citep{tilman1981,dutkiewicz2009,barton2010}. Thus, the combined NSM
  steady-state dynamics and allometric timescales (see Eq.~\eqref{t1}) predict
  that larger mammals have an intrinsic competitive advantage given a common
  resource.  

  \begin{figure}
  \centering
  \includegraphics[width=0.4\textwidth]{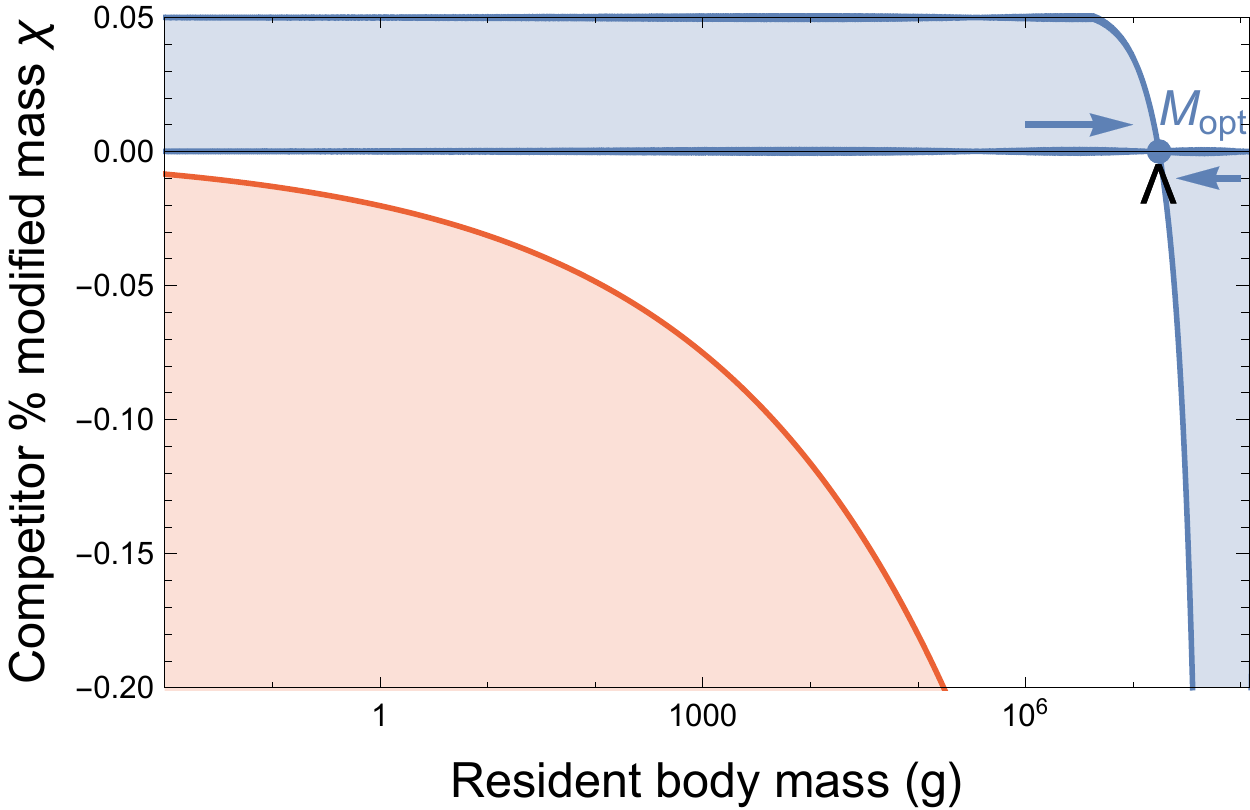}
  \caption{\small{Competitive outcomes for a resident species with body mass
      $M$ vs. a closely related competing species with modified body mass
      $M^\prime=M(1+\chi)$.  The blue region denotes proportions of modified
      mass $\chi$ resulting in exclusion of the resident species.  The red
      region denotes values of $\chi$ that result in a mass that is below the
      starvation threshold and are thus infeasible.  Arrows point to the
      predicted optimal mass from our model $M_{\rm opt}=1.748\times 10^7$,
      which may serve as an evolutionary attractor for body mass.  The black
      wedge points to the largest body mass known for terrestrial mammals
      (\emph{Deinotherium} spp.) at $1.74\times10^7$
      (g)~\citep{Smith:2010p3442}.}\label{fig:invasion}}
  \end{figure}


  However, the above resource relationships do not offer a mechanism for how
  body size is selected.  We directly assess competitive outcome between two
  closely related species: a resident species of mass $M$, and a competing
  species (denoted by $^\prime$) where individuals have a different proportion
  of body fat such that $M^\prime=M(1+\chi)$.  For $\chi < 0$, the competing
  individuals have fewer metabolic reserves than the resident species and vice
  versa for $\chi>0$.  For the allowable values of $\chi$ (see SI), the mass of
  the competitor $M'$ should exceed the minimal amount of body fat,
  $1+\chi>\epsilon_{\sigma}$, and the adjusted time to reproduce must be
  positive, which, given Eq.~\ref{t1}, implies that
  $1-\epsilon_{\lambda}^{1-\eta}\left(1+\chi\right)^{1-\eta}>0$.  These
  conditions imply that $\chi\in(-f_0M^{\gamma-1},1/\epsilon_{\lambda}-1)$
  where the upper bound approximately equals $0.05$ and the lower bound is
  mass-dependent.  The modified mass of the competitor leads to altered rates
  of starvation $\sigma(M^\prime)$, recovery $\rho(M^\prime)$, and the
  maintenance of both starving $\delta(M^\prime)$ and full consumers
  $\beta(M^\prime)$ (see the SI for derivations of competitor rates).
  Importantly, $\epsilon_\sigma$, which determines the point along the growth
  curve that defines the body composition of starved foragers, is assumed to
  remain unchanged for the competing population (see SI).

  To assess the susceptibility of the resident species to competitive
  exclusion, we determine which consumer pushes the steady-state resource
  density $R^*$ to lower values for a given value of $\chi$, with the
  expectation that a population capable of surviving on lower resource
  densities has a competitive advantage \citep{tilman1981}.  We find that for
  $M\leq 1.748\times10^7$ (g), having additional body fat ($\chi > 0$) results
  in a lower steady state resource density ($R^{\prime *}<R^*$), such that the
  competitor has an intrinsic advantage over the resident species
  (Fig. \ref{fig:invasion}).  However, for $M> 1.748\times10^7$ (g), leaner
  individuals ($\chi < 0$) have lower resource steady state densities.


  The observed switch in susceptibility as a function of $\chi$ at
  $M_{\rm opt}= 1.748\times10^7$ (g) thus serves as an attractor, such that the
  NSM predicts organismal mass to increase if $M<M_{\rm opt}$ and decrease if
  $M>M_{\rm opt}$.  This value is close to but smaller than the asymptotic
  upper bound for terrestrial mammal body size predicted by the NSM, and is remarkably close to independent estimates of the largest land
  mammals, the early Oligocene \emph{Indricotherium} at $\approx$
  $1.5\times10^7$ (g) and the late Miocene \emph{Deinotherium} at $\approx$
  $1.74\times10^7$ (g) ~\citep{Smith:2010p3442}.  Additionally, our calculation
  of $M_{\rm opt}$ as a function of mass-dependent physiological rates is
  similar to theoretical estimates of maximum body size \citep{Clauset:2009fh},
  and provides independent theoretical support for the observation of a
  `maximum body size attractor' explored by Alroy~\citep{Alroy:1998p1594}.

  An optimal size for mammals at intermediate body mass was predicted by Brown et al.\ based on reproductive maximization and the transition between hungry and full individuals \cite{Brown:1993p708}. 
  By coupling the NSM to resource dynamics as well as introducing an explicit treatment of storage, we show that species with larger body masses have an inherent competitive advantage for size classes up to $M_{\rm opt}= 1.748\times10^7$ based on resource competition. Moreover, the mass distributions in Ref. \cite{Brown:1993p708} show that intermediate mammal sizes have the greatest species diversity, in contrast to our efforts, which consider total biomass and predict a much larger $M_{\rm opt}$.
  Compellingly, recent work shows that many communities can be dominated by the biomass of the large \cite{Hempson:2015hka}.
  While the state of the environment as well as the competitive landscape will determine whether specific body sizes are selected for or against~\citep{Saarinen:2014br}, we propose that the dynamics of starvation and recovery described in the NSM provide a general selective mechanism for the evolution of larger body size among terrestrial mammals.\\

  \noindent {\bf Discussion} \\



  The energetics associated with somatic maintenance, growth, and reproduction
  are important elements that influence the dynamics of all
  populations~\citep{Stearns:1989ip}.  The NSM incorporates the dynamics of
  starvation and recovery that are expected to occur in resource-limited
  environments.  We found that incorporating allometrically-determined rates
  into the NSM predicts that: (i) extinction risk is minimized, (ii) the
  derived steady-states quantitatively reproduce Damuth's law, and (iii) the
  selective mechanism for the evolution of larger body sizes agrees with Cope's
  rule.  The NSM offers a means by which the dynamic consequences of energetic
  constraints can be assessed using macroscale interactions between and among
  species.


  \section*{Methods}
  \small{
  {\bf Analytical solution to the NSM}
  Equation~\eqref{eq:system} has three fixed points: two trivial fixed points at $(F^*,H^*,R^*)=(0,0,0)$ and $(0,0,1)$, and one non-trivial, internal fixed point at
  \begin{align}
  \label{eq:ss}
  \begin{split}
  F^* &= (\sigma-\lambda)\frac{ \alpha  \lambda  \mu ^2  (\mu +\xi  \rho )}{A (\lambda  \rho  B+\mu  \sigma  (\beta  \mu +\lambda  (\delta +\rho )))}, \\
  H^* &= (\sigma-\lambda)\frac{ \alpha  \lambda ^2 \mu  (\mu +\xi  \rho )}{A (\lambda  \rho  B+\mu  \sigma  (\beta  \mu +\lambda  (\delta +\rho )))}, \\
  R^* &= (\sigma - \lambda)\frac{\mu  }{A}.
  \end{split}
  \end{align}
  where $A=(\lambda \xi \rho +\mu \sigma )$ and
  $B=(\beta \mu \xi +\delta \lambda \xi -\lambda \mu )$. The stability of this
  fixed point is determined by the Jacobian matrix $\bf J$, with
  $J_{ij}=\partial{\dot X_i}/\partial{X_j}$, when evaluated at the internal
  fixed point, and $\mathbf{X}$ is the vector $(F,H,R)$.  The parameters in
  Eq.~\eqref{eq:system} are such that the real part of the largest eigenvalue
  of $\mathbf{J}$ is negative, so that the system is stable with respect to
  small perturbations from the fixed point.  Because this fixed point is
  unique, it is the global attractor for all population trajectories for any
  initial condition where the resource and consumer densities are both nonzero.

  {\bf Metabolic scaling relationships} The scaling relation between an
  organism's metabolic rate $B$ and its body mass $M$ at reproductive maturity
  is known to scale as $B = B_0 M^\eta$, where the scaling exponent $\eta$ is
  typically close to $2/3$ or $3/4$ for metazoans (e.g.,
  Ref.~\citep{West:2002it,Brown:2004wq}), and has taxonomic shifts for
  unicellular species between $\eta\approx 1$ in eukaryotes and
  $\eta\approx 1.76$ in bacteria \citep{DeLong:2010dy,Kempes:2012hy}.

  Several efforts have shown how a partitioning of $B$ between growth and
  maintenance purposes can be used to derive a general equation for both the
  growth trajectories and growth rates of organisms ranging from bacteria to
  metazoans
  \citep{West:2001bv,moses2008rmo,gillooly2002esa,hou,Savage:2004ed,Kempes:2012hy}. This relation is derived from the simple balance condition 
  $B_{0}m^{\eta}=E_{m}\dot{m}+B_{m}m\,,$
  \citep{West:2001bv,moses2008rmo,gillooly2002esa,hou,Savage:2004ed,Kempes:2012hy} where $E_{m}$ is the energy needed to synthesize a unit of mass, $B_{m}$ is
  the metabolic rate to support an existing unit of mass, and $m$ is the mass
  of the organism at any point in its development.  This balance has the
  general solution \citep{bettencourt,Kempes:2012hy}
  \begin{eqnarray}
  \label{m1}
  \left(\frac{m\left(t\right)}{M}\right)^{1-\eta}\!=1\!-\!\left[1\!-\!\left(\frac{m_{0}}{M}\right)^{1\!-\!\eta}\right]e^{-a\left(1\!-\!\eta\right)t/M^{1-\eta}},
  \end{eqnarray}
  where, for $\eta<1$, $M=(B_{0}/B_{m})^{1/(1-\eta)}$ is the asymptotic mass,
  $a=B_{0}/E_{m}$, and $m_0$ is mass at birth, itself varying allometrically
  (see the SI).  We now use this solution to define the timescale for
  reproduction and recovery from starvation (Fig.~\ref{fig:growth}; see
  \citep{moses2008rmo} for a detailed presentation of these timescales). The
  time that an organism takes to reach a particular mass $\epsilon M$ is given
  by the timescale
  \begin{equation}
  \label{t1}
  \tau\left(\epsilon\right) = \ln\left[\frac{1-\left(m_{0}/M\right)^{1-\eta}}{1-\epsilon^{1-\eta}}\right]\frac{M^{1-\eta}}{a\left(1-\eta\right)},
  \end{equation}
  where we define values of $\epsilon$ below to describe a variety of
  timescales, along with the rates related to $\tau$.  For example, the rate of
  reproduction is given by the timescale to go from the birth mass to the adult
  mass. The time to reproduce is given by Equation \ref{t1} as
  $t_{\lambda}=\tau\left(\epsilon_{\lambda}\right)$, where $\epsilon_{\lambda}$
  is the fraction of the asymptotic mass where an organism is reproductively
  mature and should be close to one (typically
  $\epsilon_{\lambda}\approx0.95\;$ \citep{West:2001bv}). Our reproductive
  rate, $\lambda$, is a specific rate, or the number of offspring produced per
  time per individual, defined as $\dot{F} = \lambda F$. In isolation this
  functional form gives the population growth
  $F\left(t\right) = F_{0}e^{\lambda t}$ which can be related to the
  reproductive timescale by assuming that when $t=t_{\lambda}$ it is also the
  case that $F=\nu F_{0}$, where $\nu-1$ is the number of offspring produced
  per reproductive cycle. Following this relationship the growth rate is given
  by $\lambda=\ln\left(\nu\right)/t_{\lambda}$, which is the standard
  relationship (e.g.,~\cite{Savage:2004ed}) and will scales as
  $\lambda\propto M^{\eta-1}$ for $M\gg m_{0}$ for any constant value of
  $\epsilon_{\lambda}$
  \citep{West:2001bv,moses2008rmo,gillooly2002esa,hou,Kempes:2012hy}.

  The rate of recovery $\rho = 1/t_\rho$ requires that an organism accrues
  sufficient tissue to transition from the hungry to the full state.  Since
  only certain tissues can be digested for energy (for example the brain cannot
  be degraded to fuel metabolism), we define the rates for starvation, death,
  and recovery by the timescales required to reach, or return from, specific
  fractions of the replete-state mass (see the SI, Table I, for
  parameterizations).  We define $m_{\sigma}=\epsilon_{\sigma} M$, where
  $\epsilon_{\sigma}<1$ is the fraction of replete-state mass where
  reproduction ceases. This fraction will deviate from a constant if tissue
  composition systematically scales with adult mass.  For example, making use
  of the observation that body fat in mammals scales with overall body size
  according to $M_{\rm fat}=f_{0}M^{\gamma}$ and assuming that once this mass
  is fully digested the organism starves, this would imply that
  $\epsilon_{\sigma}=1-f_{0}M^{\gamma}/M$. It follows that the recovery
  timescale, $t_{\rho}$, is the time to go from mass
  $m=\epsilon_{\sigma} \epsilon_{\lambda} M$ to $m=\epsilon_{\lambda}M$
  (Fig. \ref{fig:growth}). Using Eqs.~\eqref{m1} and \eqref{t1} this timescale
  is given by simply considering the growth curve starting from a mass of
  $m_{0}^{\prime}=\epsilon_{\sigma}\epsilon_{\lambda}M$, in which case
  \begin{eqnarray}
  \label{rhotimescale}
  t_{\rho}=\ln\left[\frac{1-\left(\epsilon_{\sigma}\epsilon_{\lambda}\right)^{1-\eta}}{1-\epsilon_\lambda^{1-\eta}}\right]\frac{M^{1-\eta}}{a^{\prime}\left(1-\eta\right)}
  \end{eqnarray}
  where $a^{\prime}=B_{0}/E_{m}^{\prime}$ accounts for possible deviations in
  the biosynthetic energetics during recovery (see the SI). It should be noted that more complicated ontogenetic models explicitly handle
  storage \citep{hou}, whereas this feature is implicitly covered by the body
  fat scaling in our framework.

  To determine the starvation rate, $\sigma$, we are interested in the time
  required for an organism to go from a mature adult that reproduces at rate
  $\lambda$, to a reduced-mass hungry state where reproduction is impossible.
  For starving individuals we assume that an organism must meet its maintenance
  requirements by using the digestion of existing mass as the sole energy
  source.  This assumption implies the metabolic balance
  $\dot{m}E_{m}^{\prime}=-B_{m}m$ or $\dot{m}=-a^{\prime}m/M^{1-\eta}$
  where $E_{m}^{\prime}$ is the amount of energy stored in a unit of existing
  body mass, which differs from $E_{m}$, the energy required to
  synthesis a unit of biomass \citep{hou}. Given the replete mass, $M$, of an organism, the
  above energy balance prescribes the mass trajectory of a non-consuming
  organism: $m\left(t\right)=Me^{-a^{\prime}t/M^{1-\eta}}$.
  The timescale for starvation is
  given by the time it takes $m(t)$ to reach $\epsilon_{\sigma} M$, which gives
  \begin{equation}
  \label{eq:sigma}
  t_{\sigma}=-\frac{M^{1-\eta}}{a^{\prime}}\ln\left(\epsilon_{\sigma}\right).
  \end{equation}
  The starvation rate is then $\sigma=1/t_{\sigma}$, which scales with
  replete-state mass as $1/M^{1-\eta}\ln\left(1-f_{0}M^{\gamma}/M\right)$.  An important
  feature is that $\sigma$ does not have a simple scaling dependence on
  $\lambda$, which is important for the dynamics that we
  later discuss.

  The time to death should follow a similar relation, but defined by a lower
  fraction of replete-state mass, $m_{\mu}=\epsilon_{\mu} M$ where $\epsilon_\mu < \epsilon_\sigma$.
  Suppose, for example, that an organism dies once it has digested all fat and
  muscle tissues, and that muscle tissue scales with body mass according to
  $M_{\rm musc}=u_{0}M^{\zeta}$.  This gives
  $\epsilon_{\mu}=1-\left(f_{0}M^{\gamma}+u_{0}M^{\zeta}\right)/M$. Muscle
  mass has been shown to be roughly proportional to body mass~\citep{Folland:2008ij} in
  mammals and thus $\epsilon_{\mu}$ is merely $\epsilon_{\sigma}$ minus a constant. The time to go from starvation to death is the total time to reach $\epsilon_{\mu}M$ minus the time to starve, or $t_{\mu}=-M^{1-\eta}\ln\left(\epsilon_{\mu}\right)/a^{\prime}-t_{\sigma}$,
  and $\mu=1/t_{\mu}$.
  }

\def\bibfont{\footnotesize}

\end{bibunit}


%
%
\clearpage

\begin{bibunit}[unsrt]

\setcounter{table}{0}
\renewcommand{\thetable}{S\arabic{table}}%
\setcounter{figure}{0}
\renewcommand{\thefigure}{S\arabic{figure}}%

\section*{Supporting Information for ``The dynamics of starvation and recovery''}

{\bf Mechanisms of Starvation and Recovery}
To understand the dynamics of starvation, recovery, reproduction, and resource competition, our framework partitions consumers into two classes: (a) a full class that is able to reproduce and, (b) a hungry class that experiences mortality at a given rate and is unable to reproduce. For the dynamics of growth, reproduction, and resource consumption, past efforts have combined the overall metabolic rate, as dictated by body size, with a growth rate that is dependent on resource abundance and, in turn, dictates resource consumption (see Refs. \citep{Kempes:2012hy,kempes2014morphological} for a brief review of this perspective). This approach has been used to understand a range of phenomena including a derivation of ontogenetic growth curves from a partitioning of metabolism into maintenance and biosynthesis (e.g. \citep{West:2001bv,moses2008rmo,hou,Kempes:2012hy}) and predictions for the steady-state resource abundance in communities of cells \citep{kempes2014morphological}. Here we leverage these mechanisms, combined with several additional concepts, to define our Nutritional State Model (NSM).

We consider the following generalized set of explicit dynamics for starvation, recovery, reproduction, and resource growth and consumption
\begin{align}
\begin{split}
\dot{F_{d}} &= \lambda_{\text{max}} F_{d} + \rho_{\text{max}}R_{d}H_{d}/k - \sigma \left(1-\frac{R_{d}}{C}\right)F_{d},  \\
\dot{H_{d}} &= \sigma \left(1-\frac{R_{d}}{C}\right)F_{d} - \rho_{\text{max}}R_{d} H_{d}/k - \mu H_{d},  \\
\dot{R_{d}} &= \alpha R_{d}\left(1-\frac{R_{d}}{C}\right) -\\
& \left[\left(\frac{\rho_{\text{max}}R_{d}}{Y_{H}k}+P_{H}\right)H_{d}+\left(\frac{\lambda_{\text{max}}}{Y_{F}}+P_{F}\right)F_{d}\right].
\label{bigdynamics}
\end{split}
\end{align}
where each term has a mechanistic meaning that we detail below (we will denote the dimensional equations with the subscript $_{d}$ before introducing the non-dimensional form that is presented in the main text). In the above equations $Y$ represents the yield coefficient (e.g., Refs. \citep{pirt,Heijnen}) which is the quantity of resources required to build a unit of organism (gram of mammal produced per gram of resource consumed) and $P$ is the specific maintenance rate of resource consumption (g resource $\cdot$ s$^{-1}$ $\cdot$ g organism$^{-1}$). If we pick $F_{d}$ and $H_{d}$ to have units of (g organisms $\cdot$ m$^{-2}$), then all of the terms of $\dot{R_{d}}$, such as $\frac{\rho\left(R_{d}\right)}{Y}H_{d}$, have units of (g resource $\cdot$ m$^{-2}$ $\cdot$ s$^{-1}$) which are the units of net primary productivity (NPP), a natural choice for $\dot{R_{d}}$. This choice also gives $R_{d}$ as (g $\cdot$ m$^{-2}$) which is also a natural unit and is simply the biomass density. In these units $\alpha$ (s$^{-1}$) is the specific growth rate of $R_{d}$, $C$ is the carrying capacity, or maximum density, of $R_{d}$ in a particular environment, and $k$ is the half-saturation constant (half the density of resources that would lead to maximum growth).

We can formally non-dimensionalize this system by the rescaling of $F=fF_{d}$, $H=fH_{d}$, $R=qR_{d}$, $t=st_{d}$, in which case our system of equations becomes
\begin{align}
\begin{split}
&\dot{F} = \frac{1}{s}\left[\lambda_{\text{max}} F + \rho_{\text{max}}\frac{R}{qk}H - \sigma \left(1-\frac{R}{qC}\right)F\right],  \\
&\dot{H} = \frac{1}{s}\left[\sigma \left(1-\frac{R}{qC}\right)F - \rho_{\text{max}}\frac{R}{qk} H - \mu H\right],  \\
& \dot{R} = \\
&\frac{1}{s}\left[\alpha R\left(1-\frac{R}{qC}\right) -\frac{q}{f}\left[\left(\frac{\rho_{\text{max}}R}{Y_{H}kq}+P_{H}\right)H+\left(\frac{\lambda_{\text{max}}}{Y_{F}}+P_{F}\right)F\right]\right].
\end{split}
\end{align}
If we make the natural choice of $s=1$, $q=1/C$, and $f=1/Y_{H}k$, then we are left with
\begin{align}
\begin{split}
\dot{F} &= \lambda F + \xi \rho RH - \sigma \left(1-R\right)F,  \\
\dot{H} &= \sigma \left(1-R\right)F - \xi \rho RH - \mu H,  \\
\dot{R} &= \alpha R\left(1-R\right) -\left(\rho R+\delta\right)H-\beta F
\label{reduceddynamics}
\end{split}
\end{align}
where we have dropped the subscripts on $\lambda_{\text{max}}$ and $\rho_{\text{max}}$ for simplicity, and $\xi\equiv C/k$, $\delta\equiv Y_{H}kP_{H}/C$, and $\beta\equiv Y_{H}k\left(\frac{\lambda_{\text{max}}}{Y_{F}}+P_{F}\right)/C$. The above equations represent the system of equations presented in the main text.
\\

{\bf Parameter Values and Estimates}
All of the parameter values employed in our model have either been directly measured in previous studies or can be estimated from combining several previous studies. Below we outline previous measurements and simple estimates of the parameters.

Metabolic rate has been generally reported to follow an exponent close to $\eta=0.75$ (e.g., Refs. \citep{West:2001bv,moses2008rmo} and the supplement for Ref. \citep{hou}). We make this assumption in the current paper, although alternate exponents, which are known to vary between roughly $0.25$ and $1.5$ for single species \citep{moses2008rmo}, could be easily incorporated into our framework, and this variation is effectively handled by the $20\%$ variations that we consider around mean trends. The exponent not only defines several scalings in our framework, but also the value of the metabolic normalization constant, $B_{0}$, given a set of data.  For mammals the metabolic normalization constant has been reported to vary between $0.018$ (W g$^{-0.75}$) and $0.047$ (W g$^{-0.75}$; Refs. \citep{hou,West:2001bv}, where the former value represents basal metabolic rate and the latter represents the field metabolic rate. We employ the field metabolic rate for our NSM model which is appropriate for active mammals (Table 1).

An important feature of our framework is the starting size, $m_{0}$, of a mammal which adjusts the overall timescales for reproduction. This starting size is known to follow an allometric relationship with adult mass of the form $m_{0}=n_{0}M^{\upsilon}$ where estimates for the exponent range between $0.71$ and $0.94$ (see Ref. \citep{peters1986ecological} for a review). We use $m_{0}=0.097M^{0.92}$ \citep{blueweiss1978relationships} which encompasses the widest range of body sizes \citep{peters1986ecological}.

The energy to synthesize a unit of biomass, $E_{m}$, has been reported to vary between $1800$ to $9500$ (J g$^{-1}$) (e.g. Refs. \citep{West:2001bv,moses2008rmo,hou}) in mammals with a mean value across many taxonomic groups of $5,774$ (J g$^{-1}$) \citep{moses2008rmo}. The unit energy available during starvation, $E^{\prime}$, could range between $7000$ (J g$^{-1}$), the return of the total energy stored during ontogeny \citep{hou} to a biochemical upper bound of $E^{\prime}=36,000$ (J g$^{-1}$) for the energetics of palmitate \citep{stryer,hou}. For our calculations we use the measured value for bulk tissues of $7000$ which assumes that the energy stored during ontogeny is returned during starvation \citep{hou}.

For the scaling of body composition it has been shown that fat mass follows $M_{\rm fat}=f_{0}M^{\gamma}$, with measured  relationships following  $0.018M^{1.25}$ ~\citep{Dunbrack:1993ec}, $0.02M^{1.19}$ ~\citep{Lindstedt:1985hm}, and $0.026M^{1.14}$ ~\citep{Lindstedt:2002td}. We use the values from \citep{Lindstedt:1985hm} which falls in the middle of this range. Similarly, the muscle mass follows $M_{\rm musc}=u_{0}M^{\zeta}$ with $u_{0}=0.383$ and $\zeta=1.00$ ~\citep{Lindstedt:2002td}.


Typically the value of $\xi=C/k$ should roughly be $2$. The value of $\rho$, $\lambda$, $\sigma$, and $\mu$ are all simple rates (note that we have not rescaled time in our non-dimensionalization) as defined in the maintext. Given that our model considers transitions over entire stages of ontogeny or nutritional states, the value of $Y$ must represent yields integrated over entire life stages. Given an energy density of $E_{d}=18200$ (J g$^{-1}$) for grass \citep{estermann} the maintenance value is given by $P_{F}=B_{0}M^{3/4}/ME_{d}$, and the yield for a full organism will be given by $Y_{F}=ME_{d}/B_{\lambda}$ (g individual $\cdot$ g grass $^{-1}$), where $B_{\lambda}$ is the lifetime energy use for reaching maturity given by
\begin{equation}
B_{\lambda}=\int_{0}^{t_{\lambda}}B_{0}m\left(t\right)^{\eta}dt.
\end{equation}
Similarly, the maintenance resource consumption rate for hungry individuals is $P_{H}=B_{0}(\epsilon_{\sigma}M)^{3/4}/(\epsilon_{\sigma}M)E_{d}$, and the yield for hungry individuals (representing the cost on resources to return to the full state) is given by $Y_{H}=ME_{d}/B_{\rho}$ where
\begin{equation}
B_{\rho}=\int_{\tau\left(\epsilon_{\sigma}\epsilon_{\lambda}\right)}^{t_{\lambda}}B_{0}m\left(t\right)^{\eta}dt.
\end{equation}
Taken together, these relationships allow us to calculate $\rho$, $\delta$, and $\beta$.

Finally, the value of $\alpha$ can be roughly estimated by the NPP divided by the corresponding biomass densities. From the data in Ref. \citep{michaletz2014convergence} we estimate the value of $\alpha$ to range between $2.81\times10^{-10}$ (s$^{-1}$) and $2.19\times10^{-8}$ (s$^{-1}$) globally. It should be noted that the value of $\alpha$ sets the overall scale of the $F^{*}$ and $H^{*}$ steady states along with $B_{tot}$ for each type. As such, we use $\alpha$ as our fit parameter to match these steady states with the data from Damuth \citep{damuth1987interspecific}. We find that the best fit is $\alpha=9.45\times10^{-9}$ (s$^{-1}$) which compares well with the calculated range above. However, two points are important to note here: first, our framework predicts the overall scaling of $F^{*}$ and $H^{*}$ independently of $\alpha$ and this correctly matches data, and second, both the asymptotic behavior and slope of $F^{*}$ and $H^{*}$ are independent of $\alpha$, such that our prediction of the maximum mammal size does not depend on $\alpha$.
\\



 \begin{table}[h]
\caption{Parameter values for mammals}
\label{param}
    \begin{center}
    \footnotesize
     \begin{tabular}{p{3.8cm} c p{2.2cm} p{1.4cm}}
     \hline
    
     Definition & Parameter & Value & References  \\
     \hline
   Asymptotic adult mass & $M$ & (g) &  \\
   Initial mass of an organism & $m_{0}$ & (g) &  \\
   Metabolic rate scaling exponent & $\eta$ & $3/4$  &  (e.g. \citep{West:2001bv,moses2008rmo,hou}) \\
   Metabolic Normalization Constant & $B_{0}$ & $0.047$ (W g$^{-0.75}$)    & \citep{hou}  \\
   Initial mass scaling exponent & $\upsilon$ & $0.92$ &  \citep{blueweiss1978relationships,peters1986ecological} \\
   Initial mass scaling normalization constant & $n_{0}$ & $0.097$ (g$^{1-\upsilon}$) & \citep{blueweiss1978relationships,peters1986ecological}  \\   
   Fat mass scaling exponent & $\gamma$ & $1.19$ & \citep{Lindstedt:1985hm} \\
   Fat scaling normalization constant & $f_{0}$ & $0.02$ (g$^{1-\eta}$) & \citep{Lindstedt:1985hm}\\
   Muscle mass scaling exponent & $\zeta$ & $1.00$  & \citep{Lindstedt:2002td} \\
   Muscle scaling normalization constantv& $u_{0}$ & $0.38$ (g$^{1-\zeta}$)  & \citep{Lindstedt:2002td} \\
   Energy to synthesis a unit of mass & $E_{m}$ & $5774$ (J gram$^{-1}$)  &  \citep{moses2008rmo,West:2001bv,hou} \\
   Energy to synthesis a unit of mass during recovery & $E_{m}^{\prime}$ & $7000$ (J gram$^{-1}$) & \citep{stryer,hou} \\
   Specific resource growth rate & $\alpha$ & $9.45\times10^{-9}$ (s$^{-1}$) & see text  \\
   Fraction of asymptotic mass representing full state & $\epsilon_{\lambda}$ & $0.95$ & \citep{West:2001bv}  \\
   Fraction of asymptotic mass representing starving state & $\epsilon_{\sigma}$ & $1-f_{0}M^{\gamma-1}$ & see text  \\
   Fraction of asymptotic mass representing death & $\epsilon_{\mu}$ & $1-\frac{f_{0}M^{\gamma}+u_{0}M^{\zeta}}{M}$ & see text \\
   Carrying capacity (maximum density) of resources & $C$ & (g m$^{-2}$) & \\
   Half Saturation Constant & $k$ & (g m$^{-2}$) &   \\
   Normalized carrying capacity & $\xi$ & $C/k\approx2$ &   \\
   Reproductive fecundity & $\nu$ & $2$ & \citep{}  \\


   \hline
    \end{tabular}
    \end{center}
   \end{table}

{\bf Rate equations for invaders with modified body mass}
We allow an invading subset of the resident population with mass $M$ to have an altered mass $M^\prime = M(1+\chi)$ where $\chi$ varies between $\chi_{\rm min} <0$ and $\chi_{\rm max}>0$, where $\chi<0$ denotes a leaner invader and $\chi > 0$ denotes an invader with additional reserves of body fat.
Importantly, we assume that the invading and resident individuals have the same proportion of non-fat tissues.
For the allowable values of $\chi$ the adjusted mass should exceed the amount of body fat, $1+\chi>\epsilon_{\sigma}$, and the adjusted time to reproduce must be positive, which given our solution for $\tau(\epsilon)$ (see main text), implies that $1-\epsilon_{\lambda}^{1-\eta}\left(1+\chi\right)^{1-\eta}>0$.
Together these conditions imply that  $\chi\in(-f_0M^{\gamma-1},1/\epsilon_{\lambda}-1)$ where the upper bound approximately equals $0.05$.

Although the starved state of invading organisms remains unchanged, the rate of starvation from the modified full state to the starved state, the rate of recovery from the starved state to the modified full state, and the maintenance rates of both, will be different, such that $\sigma^\prime = \sigma(M^\prime)$, $\rho^\prime = \rho(M^\prime)$, $\beta^\prime = \beta(M^\prime)$, $\delta^\prime = \delta(M^\prime)$.
Rates of starvation and recovery for the invading population are easily derived by adjusting the starting or ending state before and after starvation and recovery, leading to the following timescales:

\begin{align}
t_{\sigma^\prime} &= -\frac{M^{1-\eta}}{a^{\prime}}\ln \left(\frac{\epsilon_\sigma}{\chi +1}\right), \\ \nonumber
t_{\rho^\prime} &= \ln \left(\frac{1-(\epsilon_\lambda \epsilon_\sigma)^{1/4}}{1-( \epsilon_\lambda(\chi +1))^{1/4}}\right)\frac{M^{1-\eta}}{a^{\prime}\left(1-\eta\right)}.
\end{align}

The maintenance rates for the invading population require more careful consideration.
First, we must recalculate the yields $Y$, as they must now be integrated over life stages that have also been slightly modified by the addition or subtraction of body fat reserves.
Given an energy density of $E_{d}=18200$ (J g$^{-1}$) for grass \citep{estermann} the maintenance value of the invading population is given by $P_{F}=B_{0}(1+\chi)M^{3/4}/(1+\chi)ME_{d}$, and the yield for a full organism will be given by $Y_{F}=(1+\chi)ME_{d}/B^{\prime}_{\lambda}$ (g individual $\cdot$ g grass $^{-1}$) where $B^{\prime}_{\lambda}$ is the lifetime energy use for the invading population reaching maturity given by
\begin{equation}
B^{\prime}_{\lambda}=\int_{0}^{t_{\lambda^\prime}}B_{0}m\left(t\right)^{\eta}dt.
\end{equation}
where
\begin{equation}
t_{\lambda^\prime} = \frac{M^{1-\eta} }{a(1-\eta)}\ln \left(\frac{1-(m_0/M)^{1-\eta}}{1-(\epsilon_\lambda (1+\chi))^{1-\eta}} \right).
\end{equation}
Note that we do not use this timescale to determine the reproductive rate of the invading consumer---which is assumed to remain the same as the resident population---but only to calulate the lifetime energy use.
Similarly, the maintenance for hungry individuals $P^\prime_{H}=B_{0}(\epsilon_{\sigma}(1+\chi)M)^{3/4}/(\epsilon_{\sigma}(1+\chi)M)E_{d}$ and the yield for hungry individuals (representing the cost on resources to return to the full state) is given by $Y^\prime_{H}=(1+\chi)ME_{d}/B^{\prime}_{\rho}$ where
\begin{equation}
B^{\prime}_{\rho}=\int_{\tau\left(\epsilon_{\sigma}\epsilon_{\lambda}\right)}^{t_{\lambda^\prime}}B_{0}m\left(t\right)^{\eta}dt.
\end{equation}
Finally, we can calculate the maintenance of the invaders as

\begin{align}
  \delta^\prime &= P^\prime_{H}Y^\prime_{H}/\xi \\ \nonumber
  \beta^\prime &= \left(\frac{\lambda_{\rm max}}{Y^\prime_{F}}+P^\prime_{F} \right)Y^\prime_{H}/\xi.
\end{align}

To determine whether or not the invader or resident population has an advantage, we compute $R^*(M)$ and $R^*(M^\prime=M(1+\chi))$ for values of $\chi \in (-f_0M^{\gamma-1},1/\epsilon_{\lambda}-1)$, and the invading population is assumed to have an advantage over the resident population if $R^*(M^\prime)<R^*(M)$.

{\bf Sensitivity to additional death terms}

It should be noted that our set of dynamics (Equations \ref{bigdynamics} and \ref{reduceddynamics}) could include a constant death term of the form $-d_{F}F$ and $-d_{H}H$ to represent death not directly linked to starvation. Adding terms of this form to our model would simply adjust the effective value of $\lambda$ and $\mu$, and we could rewrite Equation \ref{reduceddynamics} with $\lambda^{\prime}=\lambda-d$ and $\mu^{\prime}=\mu-d$. These substitutions would not alter the functional form of our model nor the steady-states and qualitative results, however the quantitative values could shift based on the size of $d$ relative to $\lambda$ and $\mu$. 

Survivorship has a well-known functional form which changes systematically with size (e.g. \cite{calder1984}). Typically survivorship is defined using the Gompertz curve 
\begin{equation}
F=F_{0}e^{\left(c_{0}/c_{1}\right)\left(1-e^{c_{1}t}\right)}
\label{gompertz}
\end{equation}
where the parameters have the following allometric dependencies on adult mass $c_{0}=a_{0}M^{b_{0}}$ and $c_{1}=a_{1}M^{b_{1}}$, with $a_{0}=1.88\times10^{-8}$ (s g$^{-b_{0}}$), $b_{0}=-0.56$, $a_{1}=1.45\times10^{-7}$ (s g$^{-b_{1}}$), and $b_{1}=-0.27$ (see \cite{calder1984} for a review).

\begin{figure}
\centering
\includegraphics[width=0.4\textwidth]{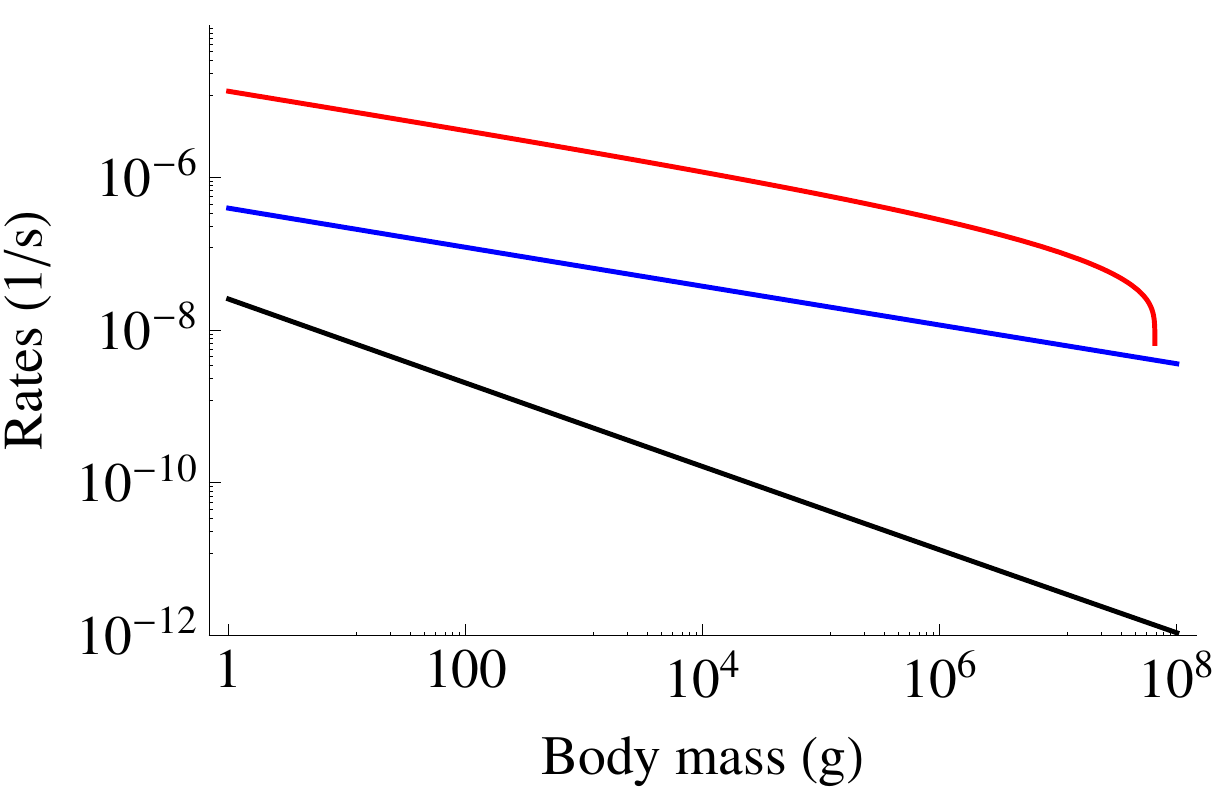}
\caption{\small{The rates of reproduction $\lambda$ (blue), starvation-based mortality $\mu$ (red), and survivorship-based death $\bar{d}$ (black) as a function of adult mass.}\label{fig:ratescomp}}
\end{figure}

We are interested in the specific death rate of the form $\dot{F}=-dF$, and using the derivative of Equation \ref{gompertz} we find that $d=c_{0}e^{c_{1}t}$. Our model considers the average rates over a population and lifecycle and the average death rate is given by 
\begin{eqnarray}
\bar{d}&=&\frac{1}{t_{\text{exp}}}\int_{0}^{t_{\text{exp}}}c_{0}e^{c_{1}t} dt \\
&=&\frac{c_{0}\left(e^{c_{1} t_{\text{exp}}}-1\right)}{c_{1}t_{\text{exp}}}
\end{eqnarray}
where $t_{\text{exp}}$ is the expected lifespan following the allometry of $t_{\text{exp}}=a_{2}M^{b_{2}}$ with $a_{2}=4.04\times10^{6}$ (s g$^{-b_{2}}$) and $b_{2}=0.30$ ~\cite{damuth1982analysis,calder1984}. Given the allometries above we have that
\begin{equation}
\bar{d}=\frac{a_{0} \left(e^{a_{1}a_{2}M^{b_{1}+b_{2}}}-1\right) M^{b_{0}-b_{1}-b_{2}}}{a_{1} a_{2}}
\end{equation}
which scales roughly like $M^{b_{0}}$ because $b_{1}$ and $b_{2}$ are close in value but opposite in sign. In Figure S\ref{fig:ratescomp} we compare the value of $\bar{d}$ to the reproductive, $\lambda$, and starvation-based mortality, $\mu$, rates. The values of $\bar{d}$ are orders of magnitude smaller than these rates for all mammalian masses, and thus, adding this non-starvation based death rate to our model does not shift our results within numerical confidence. 

\begin{figure}[h!]
\centering
\includegraphics[width=0.4\textwidth]{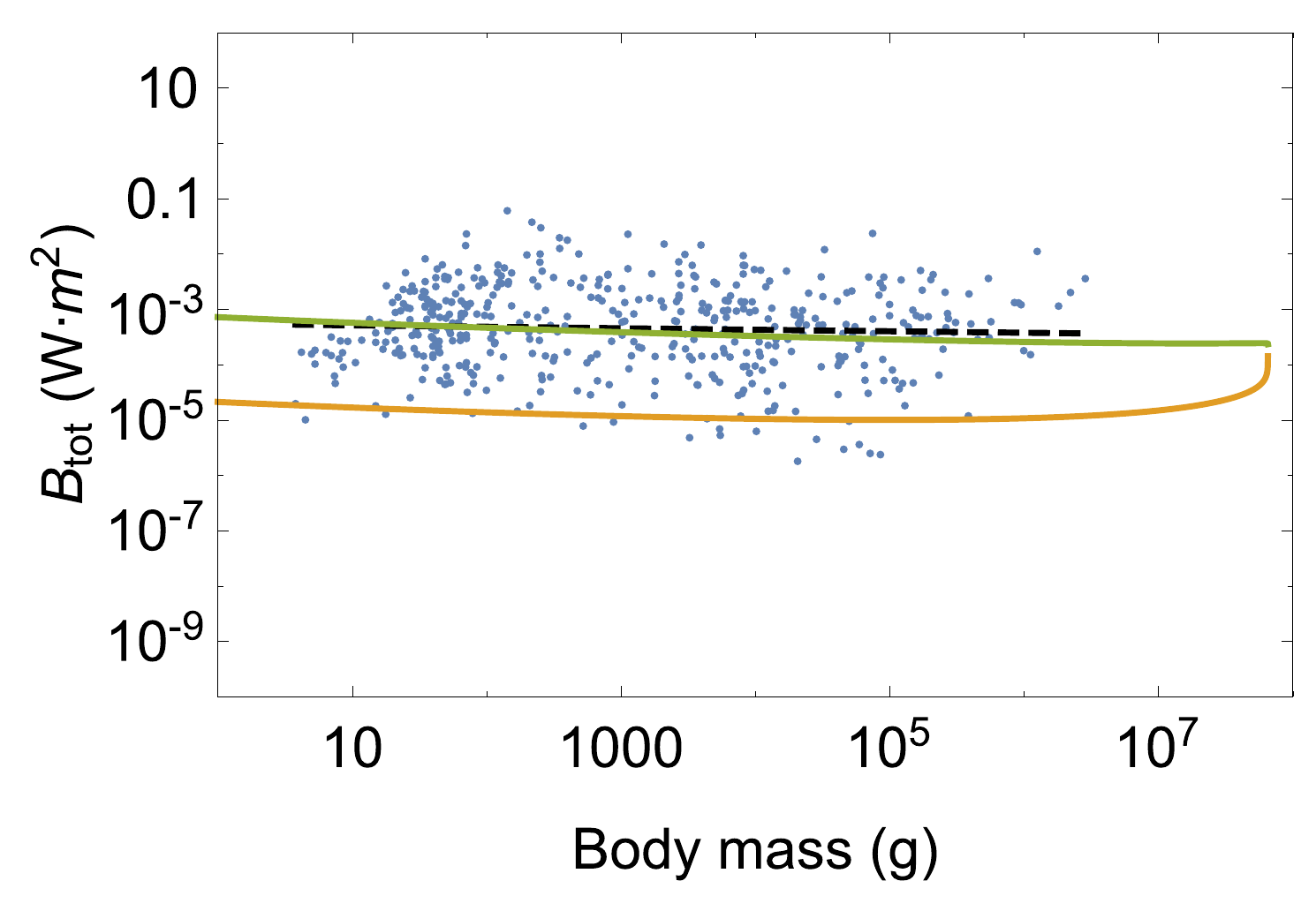}
\caption{\small{ Total energetic use $B_{\rm tot}$ of consumer populations at the steady state as a function of body mass ($F^*$ is shown in green and $H^*$ in orange).  The data are from Damuth \citep{Damuth:1987kr} and have been converted to
  total population metabolism using the allometric relationships for
  metabolic rate (e.g. Refs.~\citep{West:2001bv,hou,moses2008rmo}).}\label{fig:equivalence}}
\end{figure}

{\bf NSM and the energy equivalence hypothesis}

The energy equivalence hypothesis is based on the observation that if one assumes that the total metabolism of an ecosystem $B_{\rm tot}$ is equally partitioned between all species ($B_{i}$, the total metabolism of one species, is a constant), then the abundances should follow $N\left(M\right)B\left(M\right)=B_{i}$ implying that $N\left(M\right)\propto M^{-\eta}$, where $\eta$ is the metabolic scaling exponent \citep{allen2002,enquist1998}. As $\eta \approx 3/4$ this hypothesis is consistent with Damuth's law \citep{allen2002}. However, the actual equivalence of energy usage of diverse species has not been measured at the population level for a variety of whole populations. Figure S\ref{fig:equivalence} recasts the results of the NSM in terms of this hypothesis and shows that $F^{*}B$ is nearly constant over the same range of mammalian sizes up to the asymptotic behavior for the largest terrestrial mammals.

{\bf Application of NSM limits to aquatic mammals}
A theoretical upper bound on mammalian body size is given by $\epsilon_\sigma=0$, where mammals are entirely composed of metabolic reserves, and this occurs at $M=8.3\times 10^8$ (g), or $120$ times the mass of a male African elephant. We note this particular limit as it may have future relevance to considerations of the ultimate constraints on aquatic mammals.


\def\bibfont{\footnotesize}


\end{bibunit}

\end{document}